\documentclass[a4paper]{article}
\usepackage{authblk}
\usepackage{enumerate,amsmath,amssymb,bm,ascmac,amsthm}
\usepackage{color}
\usepackage{comment}
\usepackage{tikz}
\usetikzlibrary{calc}
\usetikzlibrary{plotmarks}

\numberwithin{equation}{section}

\usepackage{a4wide}

\usepackage{cases}
\usepackage{mathrsfs}

{\theoremstyle{plain}%
  \newtheorem{thm}{\bf Theorem}[section]%
  \newtheorem{lem}[thm]{\bf Lemma}%
  \newtheorem{prop}[thm]{\bf Proposition}%
  \newtheorem{cor}[thm]{\bf Corollary}%
}
{\theoremstyle{remark}
  \newtheorem{rem}[thm]{\bf Remark}
}
{\theoremstyle{definition}
  \newtheorem{ex}[thm]{\bf Example}
  \newtheorem{df}[thm]{\bf Definition}
}

\newcommand{\dC}{{\mathbb{C}}}%

\newcommand{\cA}{{\mathcal{A}}}
\newcommand{\cU}{{\mathcal{U}}}

\newcommand{\bu}{{\mathbf{u}}}
\newcommand{\bv}{{\mathbf{v}}}
\newcommand{\bj}{{\mathbf{j}}}
\newcommand{\bvarphi}{{\boldsymbol{\varphi}}}
\newcommand{\bbeta}{{\boldsymbol{\beta}}}
\newcommand{\bx}{{\mathbf{x}}}
\newcommand{\by}{{\mathbf{y}}}
\newcommand{\bff}{{\mathbf{f}}}
\newcommand{\bg}{{\mathbf{g}}}
\newcommand{\bh}{{\mathbf{h}}}

\DeclareMathOperator{\tr}{tr}

\title{A quantum searching model finding one of the edges of a subgraph in a complete graph}

\author[1]{Yusuke YOSHIE\footnote{E-mail: y-yoshie@ishikawa-nct.ac.jp}}

\author[2]{Kiyoto YOSHINO\footnote{E-mail: kiyoto.yosino.r2@dc.tohoku.ac.jp}}
\affil[1]{General Education, Ishikawa College, National Institute of Technology,
Ishikawa 929-0392, Japan}
\affil[2]{Graduate School of Information Sciences, Tohoku University, 
Sendai 980-8579, Japan}
\date{}
\begin {document}
\maketitle
\begin{abstract}
Some of the quantum searching models have been given by perturbed quantum walks.
Driving some perturbed quantum walks, we may quickly find one of the targets with high probability. 
In this paper, we address a discrete-time quantum walk.
We construct a quantum searching model finding one of the edges of a given subgraph in a complete graph. 
How to construct our model is that we label the arcs by $+1$ or $-1$, and define a perturbed quantum walk by the sign function on the set of arcs.
After that, we detect one of the edges labeled $-1$ by the induced sign function as fast as possible. 
This idea was firstly proposed by Segawa \textit{et~al.} in 2021. 
They only addressed the case where the subgraph forms a matching,
and obtained by a combinatorial argument that the time of finding one of the edges of the subgraph is quadratically faster than a classical searching model.
In this paper, we show that the model is valid for any subgraph, i.e., we obtain by spectral analysis a quadratic speed-up for finding one of the edges of the subgraph in a complete graph.
 \end{abstract}
\section{Introduction}
A quantum walk was introduced as a quantum analogy of a classical random walk~\cite{Gudder, Meyer}.
The origin of one of the quantum walks is seen in~\cite{Feynman}. 
In the early 1990s, Aharonov {\it et~al.}~\cite{Aharonov} reformulated it as a quantum random walk and designed a system finding an excited state as fast as possible. 
After that, research on quantum walks has been vigorous.  
Especially, these were expected to design an efficient searching system, which is called a quantum search algorithm. 
Until today, such systems have been successively proposed. 
The most remarkable algorithm is the one proposed by Grover~\cite{Grover}. 
The algorithm called Grover's algorithm realized a system detecting a target in an unordered database of $N$ items with $O(\sqrt{N})$ times, which gives a quadratic speed-up over a classical search algorithm. 
The algorithm is regarded as a search on a complete graph. This work focused on searching for a single target. As a generalization of Grover's algorithm, Boyer {\it et~al.} \cite{Boyer} proposed a searching algorithm finding one of two or more targets.
As Shenvi {\it et~al.}~\cite{Shenvi} proposed a quantum search algorithm on a hypercube, studies on quantum search on general graphs have been in the limelight. 
Moreover, Szegedy~\cite{Sze04} designed a quantum walk called a bipartite walk on a bipartite graph and gave a fundamental idea of quantum searches on graphs. 
Ambainis~{\it et al.}~\cite{Amb3} studied a quantum search on a finite grid of size $N$ with more than $2$ dimensions, which detects a target with $O(\sqrt{N})$ times. 
Besides this, quantum searching algorithms on some classes of graphs have been studied, e.g., triangular lattices~\cite{Abal}, highly symmetric graphs~\cite{Reitzner}, simplicial complexes~\cite{Matsue18} and so forth. 
Furthermore, element distinctness~\cite{Amb2} and the finding triangle problem~\cite{Magniez} were proposed. 
In these works, quantum walks often help us to detect a target efficiently.   
A quantum search algorithm is often designed by driving a perturbed quantum walk. 
It enables us to detect a target, say marked one, as fast as possible. The perturbed quantum walk is given by a time evolution operator with a perturbation on the targets. 
A time evolution operator is constructed by a product of two unitary operators called a shift operator and a coin operator~\cite{KPSS}.  
The perturbation is often given in the coin operator. 
For example, the perturbed coin operator is constructed so that it acts as the Grover coin in non-marked vertices and $-I$ in marked vertices, where $I$ is the identity operator~\cite{Amb3}. 
This difference often gives us a considerable speed-up of a quantum search.

Also, another particularly well-known algorithm was proposed by Shor~\cite{Shor}, which efficiently factors numbers.
The development of quantum computers has been actively made for executing algorithms such as those described above.
As reviewed by Huang, Wu, Fan and Zhu in~\cite{HWFZ},
tremendous advances have been made for constructing large-scale quantum computers over the last two decades,
and experimental efforts continue.
In 2019, the demonstration of quantum supremacy was first achieved using $53$ superconducting qubits~\cite{Arute}.
However, quantum computer devices are currently still small scale, and their capabilities have not reached the level beyond small demonstration algorithms.

In this paper, we aim to detect one of some edges on a complete graph on $n+1$ vertices by a perturbed quantum walk. As found in \cite{Boyer}, we prepare one or more targets and give them a perturbation. 
An idea to attach the perturbation is given by a signed graph.  
A sign function is a function from the edge set to $\{ \pm 1\}$. 
As is seen in~\cite{Harary}, the signed graph is introduced as a model for a social network. 
Our searching model begins with specifying a set of edges and labeling them as $-1$. 
Then we construct the time evolution operator of a quantum walk by the sign function.
After that, we drive the quantum walk and estimate the number of times to apply the time evolution operator until the finding probability of the negatively signed edges is sufficiently high. 
In other words, we use the perturbed quantum walk to find one of the edges of a subgraph whose edges are labeled by $-1$, say $\Gamma$, as fast as possible. 
This idea is firstly introduced by Segawa \textit{ et~al.}~\cite{SY}. In this work, the perturbed quantum walk realizes a quadratic speed-up in the case where the set of negatively signed edges is a matching in  a complete graph.  
What we would like to do now is to extend the previous result. 
More precisely, we design the quantum walk on a signed complete graph $G$ where the set of negatively signed edges forms a general graph. 
We show that the idea as in~\cite{SY} is valid for any subgraph $\Gamma$. 
Specifically, we prove that the time complexities of our quantum search and a classical search based on a random walk are as follows:
\[
\begin{cases}
O\left(n/\sqrt{|E(\Gamma)|}\right), & \text{quantum search},\\
O\left(n^2/|E(\Gamma)| \right), & \text{classical search}.
\end{cases}
\]
Thus, our model enables us to obtain a quadratic speed-up over a classical searching model for any subgraph $\Gamma$. 
We remark that the condition 
\begin{align*}
    \frac{|V(\Gamma)|}{|V(G)|}+\frac{|E(\Gamma)|}{|E(G)|} < c
\end{align*}
for some small positive constant $c$ is assumed in the main result (see Corollary~\ref{cor:main}).
This does not limit the applicability of our model and does not change the order of the searching time because we can embed the complete graph $G$ into a sufficiently large complete graph $G'$ and detect one of the edges of $\Gamma$ in $G'$.

This paper is organized as follows: In Section 2, we lay out frameworks of graphs and a sign function. In addition, we design our perturbed quantum walk by a sign function. 
In Section~3, we address matrix analysis, and estimate eigenvalues and eigenvectors of matrices which play an important role in this paper. 
Sections~4 and~5 compare the quantum searching time and classical one.
We first establish the time complexity of our quantum searching model by spectral analysis in Section~4,
and next compute the classical one in a line graph in Section~5.
Lastly, we summarize our result and draw future directions of our work in Section~6. 

\section{Preliminaries}
\label{preliminary}
\subsection{Graph and sign}
Throughout this paper, all the graphs are simple graphs, which have no loops and multiple edges.
Let $G$ be a graph.
Let $V(G)$ denote the set of vertices, and $E(G)$ the set of edges of $G$.
Write $uv$ for the edge $\{ u,v \}$, 
and $\deg_G v$ for the degree of a vertex $v$ in $G$.
Define $\cA(G):=\{ (u,v) \mid uv \in E(G)\}$, which is the set of symmetric arcs of $G$. 
For $a \in \cA(G)$, $t(a)$ and $o(a)$ denote the \emph{terminus} and \emph{origin} of $a$, respectively.
In addition, $a^{-1}$ denotes the \emph{inverse arc} of $a$. 
Namely, $t((u,v))=v$, $o((u,v))=u$ and $(u,v)^{-1}=(v,u)$.

We write the \emph{adjacency matrix} and the \emph{degree matrix} of a graph $G$ as $A(G)$ and $D(G)$, respectively.  
Let $N(G)$ be a matrix, whose rows are indexed by $V(G)$ and columns are indexed by $E(G)$, satisfying
\[
	N(G)_{v, e}=
	\begin{cases}
		1, & \text{$v \in e$},\\
		0, & \text{otherwise}.
	\end{cases}
\]
This matrix is called the \emph{incident matrix} of $G$. 
The following fact for the incident matrix is well-known:
\begin{equation}
N(G)N(G)^{\top}=A(G)+D(G). 
\label{incident1}
\end{equation}
Furthermore, let $L(G)$ denote the line graph of $G$, and then 
\begin{align*}
	N(G)^\top N(G) = A(L(G)) + 2I
\end{align*}
holds.
Here, the symbol $I$ denotes the identity matrix.
Also the symbols $J$ and $O$ denote the all-ones matrix and the all-zeros matrix, respectively.
If the size of each matrix is not clear, then we will indicate its size by a subscript.
In addition, the symbol $\bj$ denotes the normalized all-ones (column) vector.
Similarly, we write $\bj_I$ for the normalized all-ones vector indexed by a set $I$ if necessary.

For a symmetric real matrix $X$ of order $n$, we denote by $\lambda_1(X) \geq \lambda_2(X) \geq \cdots \geq \lambda_n(X)$ the eigenvalues of $X$, and write $\lambda_{\max}(X) := \lambda_1(X)$ and $\lambda_{\min}(X) := \lambda_n(X)$.
Moreover, denote by $\mathrm{Spec}(X)$ the multiset of eigenvalues of a matrix $X$.

Throughout this paper, we will use the notations introduced in the following definition. 
\begin{df}  \label{df:assumption1}
	Let $G$ be a graph.
	Let $\sigma: \cA(G) \to \{ \pm 1 \}$ be a sign function on $\cA(G)$ such that $\sigma(a^{-1})=1$ whenever $\sigma(a)=-1$ for $a \in \cA(G)$.
	In addition, we give a sign function $\tau: E(G) \to \{ \pm 1\}$ by
	\[ 
		\tau(uv)=\sigma((u,v))\cdot \sigma((v,u)).
	\]  
	If an edge $e \in E(G)$ satisfies $\tau(e)=-1$, then we call it a \emph{marked edge}.
\end{df}

\subsection{Time evolution operator}
In this subsection, we construct the time evolution operator of a perturbed quantum walk from the sign function $\sigma$ on $\cA(G)$ in Definition~\ref{df:assumption1}.
For short, we write $\deg v$ for the degree $\deg_G v$ of a vertex $v$ of $G$.
First, let us define a matrix $S$ indexed by $\cA(G)$ by
\[
S_{a,b}=
\begin{cases}
1, & a=b^{-1},\\
0, & \text{otherwise}.
\end{cases}
\]
Note that $S^{2}=I$. In addition, we give a matrix $d_{\sigma}$, whose rows are indexed by $V(G)$ and columns are indexed by $\cA(G)$, by
\[
(d_{\sigma})_{v,a}=
\begin{cases}
\frac{\sigma(a)}{\sqrt{\deg{t(a)}}}, & t(a)=v,\\
0, & \text{otherwise}.
\end{cases}
\]
It follows immediately that 
\[
(d^{*}_{\sigma})_{a,v}=
\begin{cases}
\frac{\sigma(a)}{\sqrt{\deg{t(a)}}}, & t(a)=v,\\
0, & \text{otherwise}.
\end{cases}
\]
Then it is easily checked that $d_{\sigma}d^{*}_{\sigma}=I$. We define the time evolution operator of the quantum walk by
\[ U_{\sigma}:=S(2d^{*}_{\sigma}d_{\sigma}-I),\]
whose entry is
\begin{equation*}
    (U_{\sigma})_{a,b}=
    \begin{cases}
        \frac{2\sigma(a^{-1})\sigma(b)}{\sqrt{\deg{o(a)}\deg{t(b)}}} - \delta_{a^{-1},b}, & t(b) = o(a),\\
        0, & \text{otherwise}.
    \end{cases}
\end{equation*}

Here $\delta$ is the Kronecker delta.
Let $\varphi_{t}$ be the quantum state at time $t$. Then $\varphi_{t}$ is given by
\[\varphi_{t}=U^{t}_{\sigma}\varphi_{0}. \]

Define $T_{\sigma}=d_{\sigma}Sd^{*}_{\sigma}$. 
It is checked that $T_{\sigma}$ is a matrix indexed by $V(G)$ whose entry is
\begin{equation}
(T_{\sigma})_{u,v}=
\begin{cases}
\frac{\tau(uv)}{\sqrt{\deg{u}\deg{v}}}, & uv \in E(G),\\
0, & \text{otherwise}.
\end{cases}
\label{def of T}
\end{equation}
We remark that $T_{\sigma}$ is a diagonalizable matrix since it is symmetric.

\begin{ex}  \label{ex}
We give an example in the case where $G=K_{5}$ with $V(G)=\{v_{1}, v_{2}, \dots, v_{5}\}$ and the set of marked edges is $\{ v_1v_2, v_2v_3, v_3v_4 \}$.
This graph is written in Figure~\ref{ex of model}, where the dashed edges are marked ones.

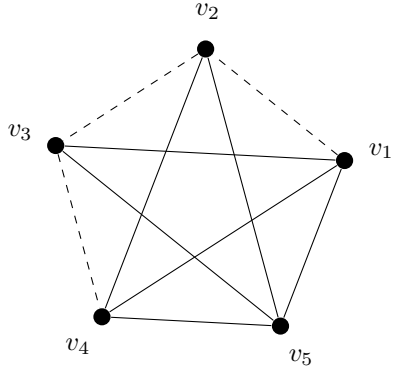
\begin{figure}[htpb]
\begin{center}
\begin{tikzpicture}
[scale = 1.0,
slim/.style={circle,fill=black, inner sep = 0.8mm},
]
\foreach \angle /\label in
{1,2,..., 5}
{
\node[slim] (v\angle) at (72*\angle+15-72:2cm) {};
\node at (72*\angle-72+15:2.5cm) {$v_{\angle}$};
}
\draw[dashed] (v1)--(v2);
\draw[dashed] (v2) to (v3);
\draw[dashed] (v3) to  (v4);
\draw (v4) to (v5);
\draw (v5) to (v1);
\draw (v1) to (v3);
\draw (v1) to (v4);
\draw (v2) to (v4);
\draw (v2) to (v5);
\draw (v3) to (v5);

\end{tikzpicture}
\caption{The graph in Example~\ref{ex}}
\label{ex of model}
\end{center}
\end{figure}

We choose a sign function $\sigma : \mathcal{A}(G) \to \mathbb{R}$ such that for $a \in \cA(G)$,
\[
\sigma(a)=
\begin{cases}
-1, & a \in \{(v_{1}, v_{2}), (v_{2}, v_{3}), (v_{3}, v_{4})\},\\
1, & \text{otherwise}.
\end{cases}
\]
Then we have
\[ 
T_{\sigma}=\frac{1}{4}\left(
\begin{array}{ccccc}
0 & -1 & 1 & 1 &1\\
-1 & 0 & -1 & 1 &1\\
1 & -1 & 0 & -1 &1\\
1 & 1 & -1 & 0 & 1 \\
1 & 1 & 1 & 1 & 0
\end{array}
\right).
\]
In addition, the entries of $U_{\sigma}$ are given by
\[
(U_{\sigma})_{a,b}=
\begin{cases}
\frac{\sigma(a^{-1})\sigma(b)}{2}-\delta_{a^{-1},b}, & \text{$t(b)=o(a)$}, \\
0, & \text{otherwise}. 
\end{cases}
\]
In particular, 
\begin{align*}
    (U_\sigma)_{(v_2,v_3),(v_3,v_2)} &= -1/2, &&(U_\sigma)_{(v_2,v_3),(v_1,v_2)} = -1/2,\\
    (U_\sigma)_{(v_2,v_3),(v_5,v_2)} &= 1/2,
    &&(U_\sigma)_{(v_2,v_3),(v_5,v_3)} = 0.
\end{align*}

\end{ex}

Here, it is known that a part of the spectrum of $U_{\sigma}$ is expressed in terms of that of $T_{\sigma}$. 
\begin{thm}[\cite{HKSS14}]
Let $U_{\sigma}$ and $T_{\sigma}$ be defined as in the above. Then it holds that 
\[ \{ e^{\pm i\theta_{\lambda}} \mid \lambda \in \mathrm{Spec}(T_{\sigma})\backslash\{ \pm 1\} \} \subset \mathrm{Spec}(U_{\sigma}),\]
where $\theta_{\lambda}=\arccos(\lambda)$.  
In addition, each unit eigenvector of $U_{\sigma}$ associated to $e^{\pm i \theta_{\lambda}}$ is given by 
\begin{equation}
    \bvarphi_{\pm\lambda}
    =\frac{1}{\sqrt{2}|\sin{\theta_{\lambda}}|}\left(d^{*}_{\sigma}-e^{\pm i\theta_{\lambda}}Sd^{*}_{\sigma}\right)\bff,  \label{eig func}
\end{equation}
where $\bff$ is a unit eigenvector of $T_{\sigma}$ associated to $\lambda$. 
That is, for $a \in \cA(G)$,
\begin{equation*}
    (\bvarphi_{\pm\lambda})_a
    =\frac{1}{\sqrt{2}|\sin{\theta_{\lambda}}|}
    \left( 
        \frac{\sigma(a)}{\sqrt{\deg{t(a)}}} \cdot \bff_{t(a)}
        -e^{\pm i\theta_{\lambda}}\frac{\sigma(a^{-1})}{\sqrt{\deg{o(a)}}} \cdot \bff_{o(a)}
    \right).
\end{equation*} 
\label{smt}
\end{thm}

\subsection{Setting and matrices for a quantum walk and a random walk}
In this subsection, we give a setting used throughout this paper, and provide matrices used to analyze a quantum walk and a random walk in Section~\ref{sec:spec}.
Recall that functions $\sigma$ and $\tau$ are given in Definition~\ref{df:assumption1}.
In the following definition, we give an additional setting.
\begin{df}\label{df:assumtion2}
	Let $n$ be a positive integer at least $2$, and set $G := K_{n+1}$.
	Assume that a marked edge with respect to the sign function $\sigma$ exists.
	Let $\Gamma$ be the subgraph of $G$ whose edge set is the set of the marked edges with respect to $\sigma$, and vertex set is the set of endpoints of these edges.	
	Let $s$ be the order of $\Gamma$, and set $t:=n+1-s$.
	Write 
	\begin{align*}
		T_\Gamma := T_\sigma 
		=\frac{1}{n} \begin{pmatrix}
			J_{s}-I_{s}-2A(\Gamma) & J_{s, t} \\
			J_{t,s} & J_{t}-I_{t}
		\end{pmatrix}.
	\end{align*}
\end{df}

Our aim is to detect one of the edges of the subgraph $\Gamma$ in $G$ as fast as possible. 
As is seen in~\cite{SY}, the quantum searching time (with respect to a quantum walk introduced by Segawa \textit{ et~al.}) is quadratically faster than the classical searching time in the case where $G$ is a complete graph and $\Gamma$ is a matching.
In this paper, we address the case where $G$ is the complete graph on $n+1$ vertices and $\Gamma$ is an arbitrary subgraph satisfying that $c|E(\Gamma)| \leq |E(G)| = n(n+1)/2$ and $c|V(\Gamma)| \leq |V(G)|=n+1$ for some constant $c>0$.

To compare the quantum walk and some known random walk in Section~\ref{sec:classical searching time}, we prepare matrices for analyzing the random walk.
\begin{df}\label{df:PQ}
	Let $\Delta$ be the graph obtained from $G=K_{n+1}$ by removing all edges of $\Gamma$.
	Let $N=N(\Delta)$ be the incidence matrix of $\Delta$.
	Then define 
	\begin{align*}
		P_\Gamma := \frac{1}{2(n-1)} A(L(\Delta)) = \frac{1}{2(n-1)} ( N^\top N - 2I ),
	\end{align*}
	and
	\begin{align*}
		Q_{\Gamma} 
		&:=
		\frac{1}{2(n-1)} (NN^\top -2I)\\
		&=
		\frac{1}{2(n-1)}
		\begin{pmatrix}
			J_s+(n-3)I_s-A(\Gamma)-D(\Gamma) & J_{s,t} \\
			J_{t,s} & J_t + (n-3) I_t
		\end{pmatrix}.
	\end{align*}
\end{df}

\section{Eigenvalues and eigenvectors for a quantum walk and a random walk}
\label{sec:spec}
In this section, we provide spectral analysis for the matrices $T_\Gamma$ and $P_\Gamma$.
It turns out behavior of the quantum walk and random walk.
We treat two similar matrices $Y$ and $Z$ instead of $T_\Gamma$ and $P_\Gamma$ since it is possible to analyze $Y$ and $Z$ in a similar manner.
\begin{df}
	Define
	\begin{align*}
	    Y := n(T_\Gamma - I) =
	    \begin{pmatrix}
			J_s-(n+1)I_s-2A(\Gamma) & J_{s,t} \\
			J_{t,s} & J_t - (n+1) I_t
		\end{pmatrix}
	\end{align*}
	and 
	\begin{align*}
	    Z := 2(n-1)(Q_{\Gamma} - I) =
	    \begin{pmatrix}
			J_s-(n+1)I_s-A(\Gamma)-D(\Gamma) & J_{s,t} \\
			J_{t,s} & J_t - (n+1) I_t
		\end{pmatrix}.
	\end{align*}
\end{df}

\subsection{Estimates of eigenvalues}
\begin{lem}[{\cite[Proposition~2.10.2]{Spg}}]\label{lem:circle}
    Let $X$ be a symmetric real matrix of order $n$.
    Set $$r_i := \sum_{j \in \{1,2,\ldots,n\} \setminus \{i\}} |X_{i,j}|.$$
    Then every eigenvalue of $X$ is contained in
    \begin{align*}
        \bigcup_{i \in \{1,\ldots,n\}} [X_{ii}-r_i,X_{ii}+r_i].
    \end{align*}
\end{lem}
This implies the following upper bound.
Also the following lower bound follows from the Rayleigh quotient.
\begin{lem} \label{lem:first:lower}
    For $X \in \{ Y,Z \}$,
    \begin{align*}
        0 \geq \lambda_1(X) \geq \bj^\top X \bj = -\frac{4|E(\Gamma)|}{n+1}.
    \end{align*}
\end{lem}

The following lemma provides a better upper bound on the maximum eigenvalue than $\lambda_{\max}(\Gamma) \leq |V(\Gamma)|-1$, which immediately follows from Lemma~\ref{lem:circle}.
Other upper bounds are also known. (For example, see~\cite{Li}.)

\begin{lem}[{\cite[Theorem~1]{Yuan}}] \label{lem:Kantan}
    The maximum eigenvalue $\lambda_{\max}(A(\Gamma))$ of a graph $\Gamma$ satisfies that
    $$\lambda_{\max}(A(\Gamma)) \leq \sqrt{2|E(\Gamma)|-|V(\Gamma)|+1}.$$
\end{lem}

\begin{thm}[{\cite[Theorem~2.4.1]{Spg}}]	\label{thm:Caurant}
	Let $X$ be a real symmetric matrix of order $n$.
	For an $i$-dimensional subspace $W$ of $\mathbb{R}^n$,	
	\begin{align*}
		\lambda_{i+1}(X) \leq \max_{x \in W^\perp, \| \bx \|_2 = 1} \bx^\top X \bx
	\end{align*}
	holds.
\end{thm}

\begin{lem}\label{lem:second}
    Both $\lambda_2(Y) \leq 2 |V(\Gamma)| - (n+3)$ and $\lambda_2(Z) \leq - (n+1)$ hold.
\end{lem}
\begin{proof}
    By Theorem~\ref{thm:Caurant}, the second maximum eigenvalue $\lambda_2(Y)$ of $Y$ satisfies that
    \begin{align*}
        \lambda_2(Y) 
        &\leq
        \max_{\bx \in \bj^\perp, \|\bx\|_2=1} \bx^\top Y\bx \\
        &=
        \max_{\bx \in \bj^\perp, \|\bx\|_2=1} 
        \bx^\top 
        \begin{pmatrix}
            J_{s}-(n+1)I_{s}-2A(\Gamma) & J_{s, t}\\
            J_{t, s} & J_{t}-(n+1)I_{t}
        \end{pmatrix} 
        \bx\\
        &=
         -(n+1)+
        \max_{\bx \in j^\perp, \|\bx\|_2=1} 
        \bx^\top 
        \begin{pmatrix}
            -2A(\Gamma) & O \\ O & O
        \end{pmatrix}
        \bx 
        \\
        &\leq -(n+1)+
        \max_{\|\by\|_2=1} \by^\top (-2A(\Gamma)) \by \\
        &= -(n+1)-2 \min_{\|\by\|_2=1} \by^\top A(\Gamma)\by \\
        &= -(n+1)-2 \lambda_{\min}(A(\Gamma)).
    \end{align*}
    By Lemma~\ref{lem:circle}, we obtain
    $$2\lambda_{\min}(A(\Gamma)) \geq -2 \left(|V(\Gamma)|-1\right).$$
    Hence we have
    \begin{align*}
        \lambda_2(Y)
        \leq -(n+1)-2\lambda_{\min}(A(\Gamma)) 
        \leq 2|V(\Gamma)| - (n+3).
    \end{align*}
    This is the desired result.
    
    Next by a similar argument,
    we have
    \begin{align*}
        \lambda_2(Z) \leq
        &\max_{\bx \in \bj^\top, \|\bx\|_2=1} \bx^\top Z\bx \\
        &=
         -(n+1)+
        \max_{\bx \in \bj^\perp, \|\bx\|_2=1} 
        \bx^\top 
        \begin{pmatrix}
            -A(\Gamma)-D(\Gamma) & O \\ O & O
        \end{pmatrix}
        \bx \\
        &= -(n+1)-\lambda_{\min}(A(\Gamma)+D(\Gamma)).
    \end{align*}
    Since 
    $$\lambda_{\min}(A(\Gamma)+D(\Gamma)) \geq 0$$
    follows from Lemma~\ref{lem:circle},
    we obtain that $\lambda_2(Z) \leq -(n+1)$.
\end{proof}

Lemmas~\ref{lem:first:lower} and~\ref{lem:second} immediately imply the following.
\begin{lem} \label{lem:1>2}
    We have
    \begin{align*}
    	\lambda_1(T_\Gamma)-\lambda_2(T_\Gamma) 
	\geq 
	\frac{n+3}{n}- \frac{4|E(\Gamma)|}{n(n+1)} -\frac{2|V(\Gamma)|}{n}.
    \end{align*}
\end{lem}

If there exists a partition $V(\Gamma)=V_1 \sqcup V_2$ such that $E(\Gamma)=\{ uv : u \in V_1 \text{ and } v \in V_2 \}$,
then $\Gamma$ is called a complete bipartite graph.
The following lemma follows from the Perron--Frobenius theorem~\cite[Theorem~2.2.1]{Spg}.
This lemma gives a condition for equality to hold in  $\lambda_{\max}(T_\Gamma) \leq 1$, which follows from Lemma~\ref{lem:first:lower}.
\begin{lem} \label{lem:theta}
    The maximum eigenvalue $\lambda_{\max}(T_\Gamma)$ of $T_\Gamma$ is equal to $1$ if and only if the graph $\Gamma$ is a complete bipartite graph of order $n+1$.
\end{lem}

Since the graph $\Gamma$ is assumed to have at least one edge,
we see by the following lemma that the largest eigenvalue of $P_\Gamma$ is less than $1$.
\begin{lem}[{\cite[Proposition~3.1.2]{Spg}}]    \label{lem:P<1}
    Assume that a graph $\Delta'$ is connected and not regular.
    Then, the maximum eigenvalue of $A(\Delta')$ is less than the maximum degree of $\Delta'$.
\end{lem}

\subsection{Eigenvectors belonging to maximum eigenvalues}

\begin{lem}\label{lem:f=j:1}
    Let $X$ be a symmetric real matrix satisfying that $\lambda_1(X) > \lambda_2(X)$.
    Let $\bff$ be a unit eigenvector belonging to $\lambda_1(X)$.
    Then for any unit vector $\bv$, 
    \begin{align*}
        \langle \bv,\bff \rangle^2 \geq \frac{\bv^\top X \bv - \lambda_2(X)}{\lambda_1(X)-\lambda_2(X)}
        = 1-\frac{\lambda_1(X) - \bv^\top X \bv}{\lambda_1(X)-\lambda_2(X)}
    \end{align*}
    holds.
\end{lem}
\begin{proof}
    Let $n$ be the order of $X$.
    Write $\lambda_1 > \lambda_2 \geq \cdots \geq \lambda_n$ for the eigenvalues of $X$,
    and fix corresponding pairwise orthogonal unit eigenvectors $\bu_1=\bff,\bu_2,\ldots,\bu_n$.
    Take an arbitrary unit vector $\bv$, and write $\bv=a_1\bu_1 +\cdots+ a_n\bu_n$.
    We have
    \begin{align*}
        \bv^\top X \bv &= \sum_{i=1}^n \lambda_i a_i^2 
        \leq \lambda_1a_1^2 + \lambda_2 \sum_{i=2}^n a_i^2 
        = \lambda_1a_1^2 + \lambda_2(1-a_1^2) \\
        &= \lambda_2 + (\lambda_1-\lambda_2) a_1^2.
    \end{align*}
    This is the desired result.
\end{proof}

\begin{prop}\label{prop:f=j}
    Let $X$ be a negative semidefinite matrix with $\lambda_2(X) < 0$.
    Let $\bff$ be a unit eigenvector belonging to the maximum eigenvalue of $X$ such that $\langle \bff,\bj \rangle \geq 0$.
    Then
    \begin{align}	\label{prop:f=j:1}
    	1-\langle \bff,\bj \rangle^2 
	\leq \frac{\bj^\top X \bj}{\lambda_2(X)}.
    \end{align}
    In particular, if $X = Y$ and $n+3 > 2|V(\Gamma)|$, then     
    \begin{align*}
        \| \bff-\bj \|_2^2 \leq \frac{8|E(\Gamma)|}{(n+1)(n+3-2|V(\Gamma)|)}.
    \end{align*}
    If $X=Z$, then
    \begin{align*}
    	\| \bff-\bj \|_2^2 
	\leq \frac{8|E(\Gamma)|}{(n+1)^2}.    
    \end{align*}
\end{prop}
\begin{proof}
	First we assume that $\bj^\top X \bj - \lambda_2(X) \leq 0$.
	Then 
	\begin{align*}
		1 \leq \frac{\bj^\top X \bj}{\lambda_2(X)}
	\end{align*}
	holds by Lemma~\ref{lem:second}, and \eqref{prop:f=j:1} follows.
	
	Next we consider the other case.
	Namely, we assume that $\bj^\top X \bj - \lambda_2(X) > 0$.
	In particular, $\lambda_1(X) > \lambda_2(X)$ holds.
    By Lemmas~\ref{lem:first:lower} and~\ref{lem:f=j:1},
    we have~\eqref{prop:f=j:1} as follows.
    \begin{align*}
        1-\langle \bff, \bj \rangle^2 
        \leq
        1 - \frac{\bj^\top X \bj - \lambda_2(X)}{\lambda_1(X)-\lambda_2(X)} 
        \leq
        1 - \frac{\bj^\top X \bj - \lambda_2(X)}{0-\lambda_2(X)} 
        =
        \frac{-\bj^\top X \bj}{ -\lambda_2(X)}.
    \end{align*}
    
    Finally since
    \begin{align*}
        \| \bff-\bj \|_2^2 
        = 
        2(1-\langle \bff,\bj \rangle) \leq 2(1-\langle \bff,\bj \rangle^2) 
        \leq 
        \frac{-2\bj^\top X \bj}{ -\lambda_2(X)},
    \end{align*}
    the other desired inequalities follow from Lemma~\ref{lem:second}.
\end{proof}

\subsection{Estimates of maximum eigenvalues}

\begin{lem} \label{lem:first:upper}
    If $n+3  > 2|V(\Gamma)|$, then
    \begin{align*}
        -\left(1 - 4\sqrt{\frac{|V(\Gamma)|}{n+3-2|V(\Gamma)|}} \right) \cdot \frac{4|E(\Gamma)|}{n+1} 
        \geq \lambda_{\max}(Y).
    \end{align*}
    We have
    \begin{align*}
        -\left(1 - 4\sqrt{\frac{|V(\Gamma)|}{n+1}} \right) \cdot \frac{4|E(\Gamma)|}{n+1} 
        \geq \lambda_{\max}(Z).
    \end{align*}    
\end{lem}
\begin{proof}
    Set $X \in \{Y,Z\}$.
    Let $\bff$ be the unit eigenvector belonging to $\lambda_{\max}(X)$ such that $(\bff,\bj) \geq 0$.
    Setting $\bv := \bff-\bj$, we have
    \begin{align*}
        \lambda_{\max}(X) &= \bff^\top X \bff \\
        &=\bj^\top X \bj + 2\bj^\top X \bv + \bv^\top X \bv.
    \end{align*}
    This together with Lemma~\ref{lem:first:lower} implies that
    \begin{align*}
        \lambda_{\max}(X) \leq -\frac{4|E(\Gamma)|}{n+1} + 2\bj^\top X \bv + 0.
    \end{align*}
    Noting that $$\bj^\top Y = \bj^\top Z,$$
    we see that
    \begin{align*}
        2\bj^\top X \bv 
        &= 2\bj^\top 
        \begin{pmatrix}
			J_s-(n+1)I_s-2A(\Gamma) & J_{s,t} \\
			J_{t,s} & J_t - (n+1) I_t
		\end{pmatrix} \bv \\
		&= 
		2\bj^\top 
        \begin{pmatrix}
			-2A(\Gamma) & O \\
			O & O
		\end{pmatrix} \bv.
    \end{align*}
    Let $\bj'$ be the unit vector whose first $|V(\Gamma)|$ elements are $1/\sqrt{|V(\Gamma)|}$ and others are $0$.
    Then we have
    \begin{align*}
        2\bj^\top 
        \begin{pmatrix}
			-2A(\Gamma) & O \\
			O & O
		\end{pmatrix} \bv
		&=
		2\sqrt{\frac{|V(\Gamma)|}{n+1}} \cdot
		{\bj'}^\top 
		\begin{pmatrix}
			-2A(\Gamma) & O \\
			O & O
		\end{pmatrix} \bv \\
		&\leq
		2\sqrt{\frac{|V(\Gamma)|}{n+1}} \cdot
		\| \bj' \|_2 \cdot
		\rho(\begin{pmatrix}
			-2A(\Gamma) & O \\
			O & O
		\end{pmatrix}) \cdot \| \bv \|_2,
    \end{align*}
    where $\rho(\cdot)$ denotes the spectral radius.
    By the Perron--Frobenius theorem~\cite[Theorem~2.2.1]{Spg},
    this spectral radius equals $2\lambda_{\max}(A(\Gamma))$.
    These together with Lemma~\ref{lem:Kantan} imply that
    \begin{align*}
        2\bj^\top X \bv 
        &\leq 
        2\sqrt{\frac{|V(\Gamma)|}{n+1}} \cdot
        2\lambda_{\max}(A(\Gamma)) \cdot \|\bj'\|_2\|\bv\|_2 \\
        &\leq 2\sqrt{\frac{|V(\Gamma)|}{n+1}} \cdot
        2\sqrt{2|E(\Gamma)|} \cdot \| \bv \|_2 \\
        & \leq 
        \frac{4|E(\Gamma)|}{n+1} \cdot \sqrt{\frac{2(n+1)|V(\Gamma)|}{|E(\Gamma)|}} \cdot \|\bv\|_2.
    \end{align*}
    Thus the desired result follows from Proposition~\ref{prop:f=j}.
\end{proof}

\begin{cor} \label{cor:T}
    If $n+3  > 2|V(\Gamma)|$, then
    \begin{align*}
        1-\left(1 - 4\sqrt{\frac{|V(\Gamma)|}{n+3-2|V(\Gamma)|}} \right) \cdot \frac{4|E(\Gamma)|}{n(n+1)} 
        \geq \lambda_{\max}(T_\Gamma)
        \geq 1-\frac{4|E(\Gamma)|}{n(n+1)}.
    \end{align*}
    Furthermore,
    the unit eigenvector $\bff$ belonging to $\lambda_{\max}(T_\Gamma)$ with $\langle \bff, \bj \rangle \geq 0$ satisfies that
    \begin{align*} 
        \| \bff-\bj \|_2^2 \leq \frac{8|E(\Gamma)|}{(n+1)(n+3-2|V(\Gamma)|)}.
    \end{align*}
\end{cor}

\begin{lem} \label{lem:l1}
    Let $m$ be a positive integer.
    A vector $\bh$ of length $m$ satisfies that $m \|\bh \|_2^2 \geq \| \bh \|_1^2$ holds. 
\end{lem}

\begin{lem} \label{lem:PQ}
    The matrices $P_{\Gamma}$ and $Q_{\Gamma}$ have the same eigenvalues except for $-1/(n-1)$.
    Furthermore, the unit eigenvector $\bff$ belonging to $\lambda_{\max}(P_{\Gamma})$ with 
    $\langle \bff,\bj_{E(\Delta)} \rangle \geq 0$ 
    satisfies that
    \begin{align*}
        1-\frac{4|E(\Gamma)|}{(n+1)^2}
        \leq
        \langle \bff,\bj_{E(\Delta)} \rangle^2 \leq 1.
    \end{align*}
\end{lem}
\begin{proof}
	Since $P_\Gamma = \frac{1}{2(n-1)}( N^\top N-2I)$ and $Q_\Gamma = \frac{1}{2(n-1)}(NN^\top-2I)$ hold, 
	the matrices $P_{\Gamma}$ and $Q_{\Gamma}$ have the same eigenvalues except for $-1/(n-1)$.
	In particular, since the maximum eigenvalue of $P_{\Gamma}$ is positive by $\tr P_{\Gamma}=0$,
	the maximum eigenvalues $\lambda_{\max}(P_{\Gamma})$ and $\lambda_{\max}(Q_{\Gamma})$ coincide.

	Next we show the second claim.
	Let $\bff$ be the unit eigenvector of $P_{\Gamma}$ belonging to $\lambda_{\max}(P_{\Gamma})$ such that $\langle \bff,\bj \rangle \geq 0$.
	By Lemmas~\ref{lem:first:lower} and~\ref{lem:second}, 
	we have
	$
		\lambda_1(P_{\Gamma}) \leq 1
	$
	and 
	\begin{align}	\label{lem:PQ:2}
		\lambda_2(P_{\Gamma}) 
		\leq 
		\max\left\{ \frac{-1}{n-1},\lambda_2(Q_\Gamma) \right\}
		\leq
		\max\left\{ \frac{-1}{n-1},1-\frac{n+1}{2(n-1)} \right\}
		=
		1-\frac{n+1}{2(n-1)} < 1.
	\end{align}
	In addition, Lemma~\ref{lem:l1} implies that
	\begin{align*}
		\bj^\top P_\Gamma \bj 
		&= \frac{1}{2(n-1)} \cdot \left( \|N \bj \|_2^2-2 \right)
		\geq \frac{1}{2(n-1)} \cdot \left( \frac{4|E(\Delta)|}{|V(\Delta)|}-2 \right)	\\	
		&= \frac{1}{2(n-1)} \cdot \left( \frac{4|E(G)|-4|E(\Gamma)|}{n+1} -2 \right) \\
		&= 1 - \frac{2|E(\Gamma)|}{(n+1)(n-1)}.
	\end{align*}
	Applying Proposition~\ref{prop:f=j} with $X := P_\Gamma-I$, we see that
	\begin{align*}
		1-\langle \bff,\bj \rangle^2 
		\leq \frac{\bj^\top (P_\Gamma-I)\bj}{\lambda_2(P_\Gamma-I)}
		\leq \frac{4|E(\Gamma)|}{(n+1)^2}.
	\end{align*}
	This is the desired result.
\end{proof}

\begin{cor} \label{cor:P}
    We have
    \begin{align*}
        1-\left(1 - 4\sqrt{\frac{|V(\Gamma)|}{n+1}} \right) \cdot \frac{2|E(\Gamma)|}{(n+1)(n-1)} 
        \geq \lambda_{\max}(P_{\Gamma})
        \geq 1-\frac{2|E(\Gamma)|}{(n+1)(n-1)}.
    \end{align*} 
    Furthermore, the unit eigenvector $\bff$ belonging to $\lambda_{\max}(P_{\Gamma})$ with $\langle \bff,\bj \rangle \geq 0$ satisfies that 
    \begin{align*}
        1-\frac{4|E(\Gamma)|}{(n+1)^2}
        \leq
        \langle \bff,\bj_{E(\Delta)} \rangle^2 \leq 1.
    \end{align*}
\end{cor}

The following lemma is needed to estimate some probability.
\begin{lem} \label{lem:ratio}
	Let $\bff$ be the unit eigenvector belonging to $\lambda_{\max}(T_{\Gamma})$ with $\langle \bff,\bj \rangle \geq 0$.
	If 
	\begin{align}	\label{prop:ratio:1}
		\frac{4|E(\Gamma)|}{|E(G)|} + \frac{4|V(\Gamma)|}{|V(G)|} \leq 1
	\end{align} 
	and $66|V(\Gamma)| \leq n+3$,
	then
	\begin{align} \label{lem:ratio:1}
		\frac{1-\langle \bff,\bj \rangle^2}{1-\lambda_{\max}(T_\Gamma)}
		\leq
		\frac{16n}{n+1} \cdot \sqrt{ \frac{|V(\Gamma)|}{n+3-2|V(\Gamma)| }}.
	\end{align}	
\end{lem}
\begin{proof}
	Assume~\eqref{prop:ratio:1}.
	Then Lemma~\ref{lem:1>2} implies that 
	$$\lambda_1(T_\Gamma)-\lambda_2(T_\Gamma) \geq (n+1)/(2n)>0.$$
	Setting
	$$\delta := 4\sqrt{ \frac{|V(\Gamma)|}{n+3-2|V(\Gamma)| }},$$
	we have
	\begin{align*}
		1-\langle \bj,\bff \rangle^2 
		\leq 
		\frac{\lambda_1(T_\Gamma)-\bj^\top T_\Gamma \bj}{\lambda_1(T_\Gamma)-\lambda_2(T_\Gamma)}
		\leq
		\frac{2n}{n+1} \cdot \delta \cdot \frac{4|E(\Gamma)|}{n(n+1)},
	\end{align*}
	by Proposition~\ref{prop:f=j} and Corollary~\ref{cor:T}.
	This together with Corollary~\ref{cor:T} implies that
	\begin{align*}
		\frac{1-\langle \bff,\bj \rangle^2}{1-\lambda_{\max}(T_\Gamma)}
		\leq 
		\frac{2n}{n+1} \cdot \frac{\delta}{1-\delta}.
	\end{align*}
	By $66|V(\Gamma)| \leq n+3$, $\delta \leq 1/2$ holds, and hence the desired conclusion follows.
\end{proof}

    We remark that in Lemma~\ref{lem:ratio}, the assumption $66 |V(\Gamma)| \leq n+3$ can be improved to a weaker assumption.
    Then the estimate~\eqref{lem:ratio:1} will become worse.
	
\section{Quantum searching time}
In this section, we estimate the quantum searching time finding one of the edges of $\Gamma$.
The method of our quantum search is based on what Ambainis {\it et al.} proposed (See~\cite{Amb3, Portugal}). 
Now, we give the outline of the process. We begin with constructing vectors $\bbeta_{\pm}$. 
In addition, we define the quantum searching time $t_{f}$ as the time converting $i\bbeta_{-}$ to $-\bbeta_{+}$, that is, $U^{t_{f}}_{\sigma}(i\bbeta_{-}) \approx -\bbeta_{+}$. 
Next, we show that $-\bbeta_{+}$ and $i\bbeta_{-}$ are sufficiently close to the final state and the initial state, respectively.
After that we estimate the finding probability $FP$ on the edges of $\Gamma$ in the final state $U_{\sigma}^{t_f}\bj$, and estimate the order of $t_f$.

Throughout this section, we let $\bff$ be the unit eigenvector belonging to $\lambda_{\max}(T_\Gamma)$ of $T_\Gamma$ such that $\langle \bff,\bj \rangle \geq 0$.
Put $$\theta_{\max}=\arccos(\lambda_{\max}(T_\Gamma)).$$
We remark that $2|V(\Gamma)| < n+3$ is assumed in lemmas and theorems in this section except for Lemma~\ref{lem:beta}.
Since $|V(\Gamma)| \leq n$ follows from $2|V(\Gamma)| < n+3$ and $\theta_{\max}>0$ is assumed in Lemma~\ref{lem:beta},
we may assume in this section that $$\theta_{\max} > 0$$ by Lemma~\ref{lem:theta}.
Then by Theorem~\ref{smt}, $e^{\pm i \theta_{\max}}$ is an eigenvalue of $U_{\sigma}$,
and $\bvarphi_{\pm \lambda_{\max}}$ in~\eqref{eig func} is a unit eigenvector of $U_\sigma$ associated to $e^{\pm i \theta_{\max}}$.
\begin{df}
    Define 
    \[ 
        \bbeta_{\pm}
        :=
        \frac{1}{\sqrt{2}}(\bvarphi_{+\lambda_{\max}}\pm \bvarphi_{-\lambda_{\max}}). 
    \]
    The quantum searing time $t_{f}$ is defined as $$\left\lfloor\frac{\pi}{2\theta_{\max}} \right\rfloor.$$
    The finding probability on the edges of $\Gamma$ in    $U_\sigma^{t_f} \bj$ is given by
    \begin{align*}  
    	FP := \left\| \left(U_\sigma^{t_f} \bj \right)     \middle|_{\cA(\Gamma)} \right\|_2^2.
    \end{align*}
\end{df}

\begin{ex}
    For a positive integer $k$, we write $P_{k+1}$ for the path graph with $k$ edges.
    We write $V(P_{k+1})=\{ v_1,\ldots,v_{k+1} \}$ and  $E(P_{k+1}) = \{ v_iv_{i+1} \mid i \in \{1,\ldots,k\} \}$.
    In Figure~\ref{fig:graph}, we provide the line chart of the probability in $U_\sigma^{t} \bj$ at time $t \in \{0,\ldots,100\} $ on the edges of $\Gamma \in \{P_2,P_3,P_4\}$ among the edges of $G = K_{100}$.
    Here, we choose $\sigma : \cA(G) \to \{ \pm 1 \}$ such that
    $\sigma( a ) = -1$ if  $a \in \{(v_1, v_{2}),\ldots,(v_k, v_{k+1})\}$, and $\sigma(a)=1$ otherwise.
%

\begin{figure}[htbp]
\begin{tikzpicture}[y=.2cm, x=1.1cm,font=\sffamily]
    \centering
	\draw (0,0) -- coordinate (x axis mid) (11.4,0);
    	\draw (0,0) -- coordinate (y axis mid) (0,35);
    	\foreach \x in {0,10,...,100}
     		\draw (0.1*\x,1pt) -- (0.1*\x,-3pt)
			node[anchor=north] {$\x$};
    	\foreach \y in {0,0.5,1}
     		\draw (1pt,30*\y) -- (-3pt,30*\y) 
     			node[anchor=east] {$\y$}; 
	\node[below=0.8cm] at (x axis mid) {Time $t$};
	\node[rotate=90, above=0.8cm] at (y axis mid) {Probability};

	\draw plot[
	    xscale=0.1, yscale = 30] 
		file {One.data};
	\draw plot[
	    xscale=0.1, yscale = 30] 
		file {Two.data};
	\draw plot[
	    xscale=0.1, yscale = 30]
		file {Three.data};  
    
    \node at (0.1*100,30*0.1184944639) [right] {$\Gamma=P_2$};
    \draw[] plot[mark=x] coordinates {(0.1*55,30*0.9777)};
    \draw node[above, align=center, xshift = 6pt] at (0.1*55,30) {$0.9777$ \\ ($t_f=55$)};
        
    \node at (0.1*100,30*0.4815394122) [above right] {$\Gamma=P_3$};
    \draw[] plot[mark=x] coordinates {(0.1*39,30*0.9664)};
    \draw node[above, align=center, xshift=10pt] at (0.1*39,30) {$0.9664$ \\ ($t_f=39$)};
    
    \node at (0.1*100,30*0.9608632514) [right] {$\Gamma=P_4$};
	\draw[] plot[mark=x] coordinates {(0.1*32,30*0.9638)};
    \draw node[above, align=center, xshift=-10pt] at (0.1*32,30) {$0.9638$ \\ ($t_f=32$)};
\end{tikzpicture}
\caption{The probability at time $t \in \{0,\ldots,100\} $ on the edges of $\Gamma \in \{P_2,P_3,P_4\}$ among the edges of $K_{100}$ \label{fig:graph}}
\end{figure}
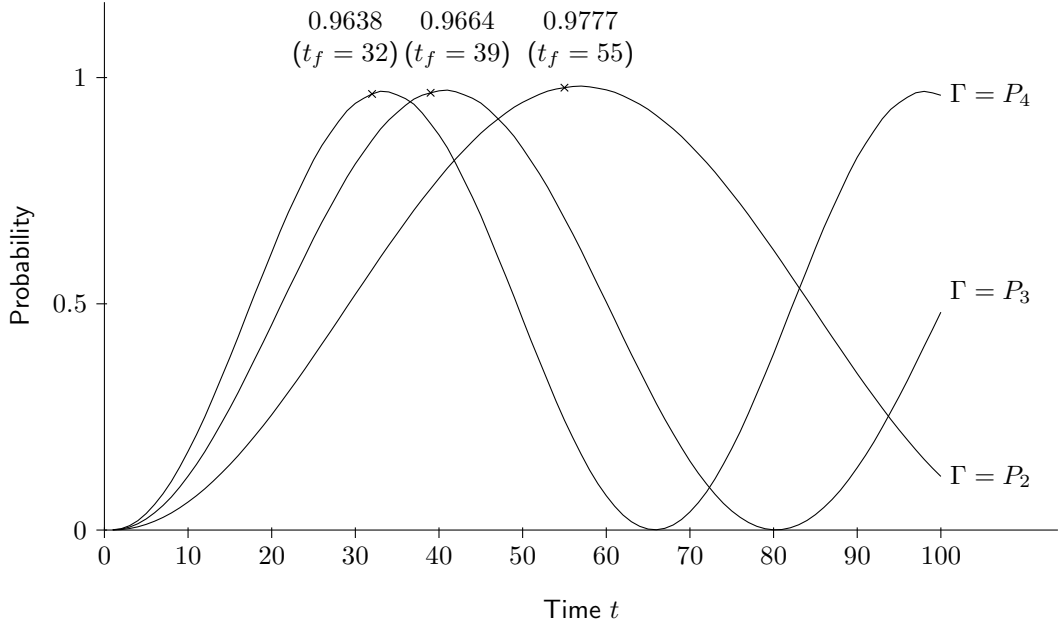
    Here, we see that the finding probabilities are high.
    In Theorem~\ref{thm:FP}, we will estimate the finding probability from below, and conclude that it converges to $1$ as $|V(\Gamma)|/|V(G)|+|E(\Gamma)|/|E(G)| \to 0$.
\end{ex}

\begin{rem}
    Assume that $\lambda_{\max} \approx 1$, or equivalently, $\theta_{\max} \approx 0$.
    In addition, we assume that $i\bbeta_-$ is close to the initial state, that is, $i\bbeta_- \approx \bj_{\cA(G)}$.
    Then we notice that
    \begin{align*}
        U_\sigma^{t_f} \bj_{\cA(G)}
        &\approx
        U_\sigma^{t_f} (i\bbeta_-) 
        =
        \frac{i}{\sqrt{2}}
        \cdot 
        U_\sigma^{t_f} (\bvarphi_{+\lambda_{\max}}- \bvarphi_{-\lambda_{\max}}) \\
        &=
        \frac{i}{\sqrt{2}} ( e^{i \theta_{\max} t_f} \bvarphi_{+\lambda_{\max}}- e^{-i \theta_{\max} t_f} \bvarphi_{-\lambda_{\max}}) \\
        &\approx
        \frac{i}{\sqrt{2}} ( e^{\frac{i \pi}{2}} \bvarphi_{+\lambda_{\max}}- e^{-\frac{i \pi}{2}} \bvarphi_{-\lambda_{\max}}) \\
        &=
        \frac{-1}{\sqrt{2}} \left( \bvarphi_{\theta_{\max}} + \bvarphi_{-\lambda_{\max}} \right) 
        = -\bbeta_+.
    \end{align*}
    This is why we define $t_f = \left\lfloor\frac{\pi}{2\theta_{\max}} \right\rfloor$.
    In fact, the assumptions will be verified below by applying Corollary~\ref{cor:T}, which is spectral analysis for $T_\Gamma$, and this observation is valid.
\end{rem}

We have the following lemma.
\begin{lem}	\label{lem:beta}
    Assume $\theta_{\max} > 0$.
	Then we have	
	\begin{align*}
		\| U^{t_{f}}_{\sigma}(i\bbeta_{-}) -  (-\bbeta_{+}) \|_2^2
		\leq
		\frac{16|E(\Gamma)|}{n(n+1)}.
	\end{align*}
\end{lem}
\begin{proof}
	We have
	\begin{align*}
		\| U^{t_{f}}_{\sigma}(i\bbeta_{-}) - (-\bbeta_{+}) \|_2^2
		&=
		\| \frac{i}{\sqrt{2}}\left( e^{i t_f \theta_{\max}} \bvarphi_{+\lambda_{\max}} - e^{-i t_f \theta_{\max}} \bvarphi_{-\lambda_{\max}} \right) 
		- \frac{-1}{\sqrt{2}} \left( \bvarphi_{+\lambda_{\max}} + \bvarphi_{-\lambda_{\max}} \right) \|_2^2 \\
		&\leq
		2 \cdot \| \frac{1}{\sqrt{2}}\left( i e^{i t_f \theta_{\max}} + 1 \right) \bvarphi_{+\lambda_{\max}} \|_2^2
		+ 2 \cdot \| \frac{1}{\sqrt{2}} \left( -ie^{-i t_f \theta_{\max}}  + 1 \right)\bvarphi_{-\lambda_{\max}} \|_2^2 \\
		&=
		2 \cdot | ie^{i t_f \theta_{\max}} +1 |^2 \\
		&=
		4( 1 - \cos \left (\frac{\pi}{2} - \left\lfloor \frac{\pi}{2 \theta_{\max}} \right\rfloor \cdot \theta_{\max} \right) ) \\
		&\leq
		4( 1 - \cos \left (\frac{\pi}{2} - \left( \frac{\pi}{2 \theta_{\max}}  -1 \right)  \cdot \theta_{\max} \right) ) \\
		&= 
		4( 1 - \cos(\theta_{\max}) ) \\
		&= 
		4( 1 - \lambda_{\max}(T_\Gamma)).
	\end{align*}
	This together with Corollary~\ref{cor:T} implies the desired conclusion.
\end{proof}

\begin{lem} \label{lem:l2}
    Let $n$ be a positive number.
    If a vector $\bh$ satisfies that $n \geq \bh_i \geq 0$ for every $i$,
    then $\|\bh \|_2^2 \leq n \| \bh \|_1$ holds. 
\end{lem}
\begin{proof}
    Under the condition that $E := \|\bh \|_1$ is constant,
    the value of $\|\bh \|_2^2$ achieves the maximum value 
    $$n^2 \left( \lfloor E/n \rfloor + (E/n-\lfloor E/n \rfloor)^2 \right)$$
    when 
    $$
        \bh = ( \overbrace{n,n,\ldots,n}^{\lfloor E/n \rfloor}, E-n \lfloor E/n \rfloor, 0,\ldots,0 )^\top.
    $$
    Since $E/n-\lfloor E/n \rfloor$ is less than $1$,
    the maximum value is bounded from above by $$n^2 \left( \lfloor E/n \rfloor + (E/n-\lfloor E/n \rfloor) \right).$$
    This is the desired result.
\end{proof}

\begin{thm}	\label{thm:beta-}
    If $2|V(\Gamma)| < n+3$, then 
    \begin{align*}
        \| i\bbeta_- - \bj_{\cA(G)} \|_2^2 
        \leq
        \frac{(12+8\sqrt{2})|E(\Gamma)|}{(n+1)(n+3-2|V(\Gamma)|)}.
    \end{align*}
\end{thm}
\begin{proof}
    Let $\bg$ be the vector indexed by $V(G)$ such that $n\sqrt{n+1} \cdot \bg_v$ is the number of arcs $a \in \cA(\Gamma)$ with $v = t(a)$ and $\sigma(a)= -1$.
    Then
    $
        d_\sigma \bj_{\cA(G)} = \bj_{V(G)} -2\bg.
    $
    Recalling
    \begin{align*}
        \bbeta_- 
        = \frac{1}{\sqrt{2}}\left(\bvarphi_{+\lambda_{\max}} - \bvarphi_{-\lambda_{\max}} \right) 
        = \frac{-i \sin \theta_{\max}}{|\sin \theta_{\max}|} \cdot S d_\sigma^* \bff
        =
        -i S d_\sigma^* \bff
        .
    \end{align*}
    we have
    \begin{align*}
        |1 - \langle \bj_{\cA(G)}, i\bbeta_- \rangle |
        &= | 1- \langle \bj_{\cA(G)} , S d_\sigma^* \bff \rangle |
        = | 1-\langle \bj_{\cA(G)} , d_\sigma^* \bff \rangle |
        = | 1-\langle \bj - 2\bg , \bff \rangle|\\
        &\leq | 1- \langle \bj, \bff \rangle | + 2 |\langle \bg , \bff-\bj \rangle| + 2 |\langle \bg , \bj \rangle|.
    \end{align*}
    We estimate the three terms below.
    First since $$\| n\sqrt{n+1} \cdot \bg \|_2 \leq \sqrt{ n |E(\Gamma)|} $$ follows from Lemma~\ref{lem:l2}, we see that
    \begin{align*}
        | \langle \bg, \bff-\bj \rangle |
        &= \frac{1}{n\sqrt{n+1}} \cdot | \langle n\sqrt{n+1} \cdot \bg, \bff-\bj \rangle | \\
        &\leq  \frac{1}{n\sqrt{n+1}} \cdot  \| n\sqrt{n+1} \cdot \bg \|_2  \cdot \| \bff-\bj \|_2 \\
        &\leq \sqrt{\frac{|E(\Gamma)|}{n(n+1)}} \cdot \| \bff-\bj \|_2.
    \end{align*}
    This together with Corollary~\ref{cor:T} implies that
    \begin{align*}
        | \langle \bg, \bff-\bj \rangle | 
        \leq \sqrt{\frac{8|E(\Gamma)|}{(n+1)(n+3-2|V(\Gamma)|)}} \cdot \sqrt{\frac{|E(\Gamma)|}{n(n+1)}}.
    \end{align*}
    Next we have
    \begin{align*}
        \langle \bg,\bj \rangle
        = \frac{|E(\Gamma)|}{n(n+1)},
    \end{align*}
    and
    \begin{align*}
        2\left| 1 - \langle \bj,\bff \rangle \right| 
        = \| \bff-\bj \|_2^2
        \leq \frac{8|E(\Gamma)|}{(n+1)(n+3-2|V(\Gamma)|)}
    \end{align*}
    by Corollary~\ref{cor:T}.
    Therefore, we obtain that
    \begin{align*}
        | 1 - \langle \bj_{\cA(G)}, i \bbeta_- \rangle |
        &\leq | 1 - \langle \bj, \bff \rangle | + 2 |\langle \bg , \bff-\bj \rangle| + 2 |\langle \bg , \bj \rangle| \\
        &\leq \frac{(6+4\sqrt{2})|E(\Gamma)|}{(n+1)(n+3-2|V(\Gamma)|)}.
    \end{align*}
    Since $\| \bj_{\cA(G)}-i\bbeta_- \|_2^2 = 2(1-\langle \bj_{\cA(G)}, i\bbeta_- \rangle)$, the desired conclusion follows.
\end{proof}

\begin{thm}	\label{thm:beta+}
    Assume~\eqref{prop:ratio:1} and $66|V(\Gamma)| \leq n+3$.
    Then
    \begin{align*}
    	\| -\bbeta_+|_{\cA(\Gamma)} \|_2^2 
	\geq 1 - \frac{2|E(\Gamma)|}{n(n+1)} - 16 \sqrt{\frac{|V(\Gamma)|}{n+3-2|V(\Gamma)|}}.
    \end{align*}
\end{thm}
\begin{proof}
	Set $FP' := \| \bbeta_+|_{\cA(\Gamma)} \|_2^2$. 
	Recall that
    \begin{align*}
        \bbeta_+ 
        = \frac{1}{\sqrt{2}}\left(\bvarphi_{+\lambda_{\max}} + \bvarphi_{-\lambda_{\max}} \right) 
        = \frac{1}{|\sin \theta_{\max}|} \cdot \left( d_\sigma^*  - \cos \theta_{\max} S d_\sigma^* \right)\bff.
    \end{align*}
    Since $D(\Gamma)-A(\Gamma)$ is positive semidefinite, by definition we have
    \begin{align*}
	 FP' 
	 &= \| \bbeta_+|_{\cA(\Gamma)} \|_2^2 \\
        &= \frac{1}{n\sin^2 \theta_{\max}} \cdot \bff^\top
            \begin{pmatrix}
                    (1+\cos^2 \theta_{\max}) D(\Gamma) + 2\cos \theta_{\max} A(\Gamma) & O \\
                    O & O
            \end{pmatrix} \bff \\
        &\geq \frac{(1+\cos \theta_{\max})^2}{n\sin^2 \theta_{\max}} \cdot \bff^\top
            \begin{pmatrix}
                    A(\Gamma) & O \\
                    O & O
            \end{pmatrix} \bff \\
        &= 
		\frac{1+\lambda_{\max}(T_\Gamma)}{2n(1-\lambda_{\max}(T_\Gamma))} \cdot \bff^\top
            \begin{pmatrix}
                    2A(\Gamma) & O \\
                    O & O
            \end{pmatrix} \bff,
    \end{align*}
    where the $(1,1)$-block is indexed by $V(\Gamma)$.
    We notice that
    \begin{align*}
        \begin{pmatrix}
                2A(\Gamma) & O \\
                O & O
        \end{pmatrix} 
        = -n\cdot T_\Gamma + J_{n+1} - I_{n+1},
    \end{align*}
    and have
    \begin{align*}
        \bff^\top \left( -n\cdot T_\Gamma + J_{n+1} - I_{n+1} \right) \bff
        &= -n \cdot \lambda_{\max}(T_\Gamma) + (n+1) \cdot \langle \bff, \bj \rangle^2 - 1 \\
        &= n \cdot (1-\lambda_{\max}(T_\Gamma)) - (n+1)( 1- \langle \bff, \bj \rangle^2). 
    \end{align*}
    Thus
    \begin{align*}
        FP'
        &\geq
        \frac{1+\lambda_{\max}(T_\Gamma)}{2} 
        \left( 1 - \frac{n+1}{n} \cdot \frac{1-\langle \bff,\bj \rangle^2}{1-\lambda_{\max}(T_\Gamma)} \right).
    \end{align*}
   By Lemma~\ref{lem:ratio}, we obtain
   \begin{align*}
        FP' 
        \geq
        \frac{1+\lambda_{\max}(T_\Gamma)}{2} 
        \left( 1 - 16 \sqrt{\frac{|V(\Gamma)|}{n+3-2|V(\Gamma)|}} \right).
    \end{align*}
    In particular, by Corollary~\ref{cor:T},
   \begin{align*}
        FP' 
        &\geq
        \left( 1 - \frac{2|E(\Gamma)|}{n(n+1)} \right) \cdot
        \left( 1 - 16 \sqrt{\frac{|V(\Gamma)|}{n+3-2|V(\Gamma)|}} \right) \\
        &\geq
        1 - \frac{2|E(\Gamma)|}{n(n+1)} - 16 \sqrt{\frac{|V(\Gamma)|}{n+3-2|V(\Gamma)|}}.
    \end{align*}
    This is the desired result.
\end{proof}

\begin{thm} \label{thm:FP}
	Assume that~\eqref{prop:ratio:1} and $66|V(\Gamma)| \leq n+3$.
	Then 
	\begin{align*}
		FP 
		\geq
		1
		- 22 \sqrt{
		    \frac{|E(\Gamma)|}{(n+1)(n+3-2|V(\Gamma)|)}
		}
		- 32 \sqrt{\frac{|V(\Gamma)|}{n+3-2|V(\Gamma)|}}.
	\end{align*}
	In particular, the finding probability $FP$ converges to $1$ as $$\frac{|V(\Gamma)|}{|V(G)|}+\frac{|E(\Gamma)|}{|E(G)|} \to 0.$$
\end{thm}
\begin{proof}
	Set $\cU := U_\sigma^{t_f}$.
	We have
	\begin{align*}
		\left\| 
		    \left(
		        \cU \bj
		    \right)
		\middle|_{\cA(\Gamma)}
		\right\|_2 
		&\geq
		\left\| -\bbeta_{+}|_{\cA(\Gamma)} \right\|_2 
		-\left\| 
		    \left(
		        \cU (i\bbeta_{-}) - (-\bbeta_{+}) 
		    \right)
		\middle|_{\cA(\Gamma)}
		\right\|_2
		-\left\| 
		    \cU 
		    \left(
		        \bj_{\cA(G)} - i\bbeta_{-} 
		    \right)
		\middle|_{\cA(\Gamma)} 
		\right\|_2 \\
		&\geq
		\left\| \bbeta_{+}|_{\cA(\Gamma)} \right\|_2^2 
		-\left\| \cU (i\bbeta_{-}) - (-\bbeta_{+}) \right\|_2
		-\left\| \bj_{\cA(G)} - i\bbeta_{-} \right\|_2.
	\end{align*}
	By Lemma~\ref{lem:beta}, Theorems~\ref{thm:beta-} and~\ref{thm:beta+}, we have
	\begin{align*}
	    &\left\| \bbeta_{+}|_{\cA(\Gamma)} \right\|_2^2 
		-\left\| \cU (i\bbeta_{-}) - (-\bbeta_{+}) \right\|_2
		-\left\| \bj_{\cA(G)} - i\bbeta_{-} \right\|_2 \\
		&\geq 
		1 - \frac{2|E(\Gamma)|}{n(n+1)} - 16 \sqrt{\frac{|V(\Gamma)|}{n+3-2|V(\Gamma)|}}
		-\sqrt{\frac{16|E(\Gamma)|}{n(n+1)}} 
		-\sqrt{\frac{(12+8\sqrt{2})|E(\Gamma)|}{(n+1)(n+3-2|V(\Gamma)|)}} \\
		&\geq
		1
		- 11 \sqrt{ \frac{|E(\Gamma)|}{(n+1)(n+3-2|V(\Gamma)|)} }
		- 16 \sqrt{\frac{|V(\Gamma)|}{n+3-2|V(\Gamma)|}}.
	\end{align*}
	Here note that $2|E(\Gamma)| \leq n(n+1)$ and hence $$\frac{2|E(\Gamma)|}{n(n+1)} \leq \sqrt{\frac{2|E(\Gamma)|}{n(n+1)}}.$$
	Recalling that 
	$FP
	=
	\left\| 
		    \left(
		        \cU \bj
		    \right)
		\middle|_{\cA(\Gamma)}
		\right\|_2^2,
	$
	we have the desired conclusion.
\end{proof}

\begin{thm}\label{thm:qtime}
    If $66|V(\Gamma)| \leq n+3$,
    then the quantum searching time $t_f$ is the order of 
    $
        n/\sqrt{|E(\Gamma)|}.
    $
\end{thm}
\begin{proof}
    Recall that the quantum searching time is the order of
    \begin{align*}
        \frac{1}{\arccos(\lambda_{\max}(T_\Gamma))}.
    \end{align*}
    Since for any $x \in (0,1)$,
    \begin{align*}
        \left| \frac{1}{\sqrt{2x}} - \frac{1}{\arccos(1-x)} \right| \leq 1
    \end{align*}
    holds, the quantum searching time is the order of $$\frac{1}{\sqrt{2(1-\lambda_{\max}(T_\Gamma))}}.$$
    By Corollary~\ref{cor:T}, we see that
    \begin{align*}
        \frac{1}{2(1-\lambda_{\max}(T_\Gamma))}
        &\leq
        \frac{1}{2 \cdot \left(1 - 4\sqrt{\frac{|V(\Gamma)|}{n+3-2|V(\Gamma)|}} \right) \cdot \frac{4|E(\Gamma)|}{n(n+1)}}.
    \end{align*}
    Noting that
    \begin{align*} 
    	4\sqrt{\frac{|V(\Gamma)|}{n+3-2|V(\Gamma)|}}
    \end{align*} 
    is at most $1/2$ if $66|V(\Gamma)| \leq n+3$,
    we derive the desired conclusion.
\end{proof}

\begin{cor} \label{cor:main}
	For any positive number $\varepsilon$, there exists a sufficiently small constant $c > 0$ such that the quantum searching time $t_f$ is the order of $n/\sqrt{|E(\Gamma)|}$
	and
	the finding probability $FP$ is at least $1-\varepsilon$
	if
	$$\frac{|V(\Gamma)|}{|V(G)|}+\frac{|E(\Gamma)|}{|E(G)|} < c.$$
\end{cor}

\section{Classical searching time} \label{sec:classical searching time}
In this section, we evaluate the classical searching time, that is, the expected value of the first hitting time to a marked edge. 
The classical search in this paper is given by an isotropic random walk on the line graph of $G=K_{n+1}$. 
We gave a transition matrix $P_\Gamma$ on $\dC^{E(G)}$ in Definition~\ref{df:PQ}. 
Then the classical searching time is given by
\begin{equation}
t_{c}= \bj_{E(\Delta)}^\top (I-P_{\Gamma})^{-1}  \bj_{E(\Delta)}
\label{hitting} 
\end{equation}
in~\cite{Sze04} if the initial state is $\bj_{E(\Delta)}$. 
Here note that the maximum eigenvalue of $P_\Gamma$ is less than $1$ by Lemma~\ref{lem:P<1}.
By using (\ref{hitting}), we estimate the order of the classical searching time in terms of spectrum. 

\begin{lem} \label{lem:random1}
    If $64 |V(\Gamma)| \leq n+1$, then 
    \begin{align*}
        \left|
            t_c - \frac{1}{1-\lambda_{\max}(P_\Gamma)}
        \right|
        \leq
        4.
    \end{align*}
\end{lem} 
\begin{proof}
    Set $e := |E(\Delta)|$.
    Let $\bff$ be the unit eigenvector of $P_\Gamma$ belonging to $\lambda_{\max}(P_{\Gamma})$ with 
    $\langle \bff,\bj_{E(\Delta)} \rangle \geq 0$.
    Fix pairwise orthogonal unit eigenvectors $\bu_1=\bff, \bu_2, \ldots, \bu_e$ of $P_\Gamma$ belonging to $\lambda_1(P_\Gamma) \geq \cdots \geq \lambda_e(P_\Gamma)$.
    Write $\bj_{E(\Delta)} = a_1 \bu_1 + \cdots + a_e \bu_e$.
    Noting that $a_i = \langle \bu_i, \bj_{E(\Delta)} \rangle$,
    we have
    \begin{align*}
        t_{c}
        = \bj_{E(\Delta)}^\top (I-P_{\Gamma})^{-1}  \bj_{E(\Delta)} 
        = \sum_{i=1}^e \frac{\langle \bu_i, \bj_{E(\Delta)} \rangle^2}{1-\lambda_i(P_\Gamma)}.
    \end{align*}
    Hence
    \begin{align*}
        \frac{\langle \bff,\bj_{E(\Delta)} \rangle^2}{1-\lambda_{\max}(P_\Gamma)}
        \leq
        t_c
        \leq
        \frac{1}{1-\lambda_{\max}(P_\Gamma)}.
    \end{align*}
    By Corollary~\ref{cor:P}, we have
    \begin{align*}
        0 
        \leq 
        \frac{1}{1-\lambda_{\max}(P_\Gamma)} - t_c
        \leq
        \frac{1 - \langle \bff,\bj_{E(\Delta)} \rangle^2}{1-\lambda_{\max}(P_\Gamma)} 
        \leq
        \frac{2(n-1)}{n+1} \cdot \frac{1}{1-4\sqrt{\frac{|V(\Gamma)|}{n+1}}}.
    \end{align*}
    Noting that 
    $$4\sqrt{\frac{|V(\Gamma)|}{n+1}} \leq 1/2$$ if $64|V(\Gamma)| \leq n+1$,
    we conclude the desired result.
\end{proof}

\begin{lem} \label{lem:random2}
    If $64 |V(\Gamma)| \leq n+1$, then 
    \begin{align*}
        \frac{(n+1)(n-1)}{|E(\Gamma)|}
        \geq
        \frac{1}{1-\lambda_{\max}(P_\Gamma)}
        \geq
        \frac{(n+1)(n-1)}{2|E(\Gamma)|}.
    \end{align*}
\end{lem}
\begin{proof}
    This follows from Corollary~\ref{cor:P}.
\end{proof}

Lemmas~\ref{lem:random1} and~\ref{lem:random2} imply the following theorem.
\begin{thm} \label{thm:classical}
    If $64 |V(\Gamma)| \leq n+1$, then
    the classical searching time $t_c$ satisfies 
    $$t_c = \Theta\left(\frac{n^2}{|E(\Gamma)|}\right).$$
\end{thm}

By Corollary~\ref{cor:main} and Theorem~\ref{thm:classical}, we conclude that our model achieves a quadratic speed-up over a classical searching model.

\section{Summary and discussion}
In this paper, we drive a quantum searching model in $G=K_{n+1}$ detecting one of the edges of a subgraph $\Gamma$ whose edges are negatively signed by a map from $E(G)$ to $\{ \pm 1\}$. 
As a result, we could find such an edge within the time complexity of $O(n/\sqrt{|E(\Gamma)|})$ while a searching model given by a classical random walk requires the time complexity of $O(n^{2}/|E(\Gamma)|)$. 
Therefore, the model realizes a quadratic speed-up over a classical searching model. 
This result is an extension of the one as in Segawa \textit{et.~al}~\cite{SY} which only treated the case where $\Gamma$ forms a matching, and shows that the model is valid for any subgraph. 

Our model only finds an edge of a specified subgraph in a complete graph. One of our future work is to reformulate this model in an arbitrary graph. In addition, we hope to construct searching models which reveal more detailed graph-structure, e.g., maximum degree, diameter and so on. Here, we constructed the model by the sign function. 
This sign function is regarded as an edge coloring of a graph. 
We believe that this model is related to some fields, e.g., graph-coloring theory, complex network and so forth. 

\section*{Acknowledgement}
The authors are grateful to Professor Munemasa for his helpful comments on spectral analysis.
K.~Yoshino is supported by JSPS KAKENHI Grant Number JP21J14427 and a scholarship from Tohoku University, Division for Interdisciplinary Advanced Research and Education.

\section*{Data availability}
The data that support the findings of this study are available from the corresponding author upon reasonable request.

\end{document}